\documentclass[journal=jctcce,manuscript=article,layout=twocolumn]{achemso}

\usepackage{amsmath,amssymb,graphicx,xcolor,tikz}
\usepackage{hyperref}
\setkeys{acs}{doi=true}
\usepackage{doi}

\usetikzlibrary{arrows.meta,backgrounds,calc,decorations.pathreplacing,fit,positioning}
\usepackage[T1]{fontenc}
\usepackage[utf8]{inputenc}
\newcommand{\abs}[1]{\left\lvert#1\right\rvert}

\definecolor{posterconv}{HTML}{FF8A7B}
\definecolor{posterconvborder}{HTML}{C83220}
\definecolor{posterhidden}{HTML}{8BB5F1}
\definecolor{posterhiddenborder}{HTML}{3260B0}
\definecolor{posterscalar}{HTML}{356A92}
\definecolor{posteroutput}{HTML}{8ED322}

\usepackage[normalem]{ulem}

\title{A Neural Density Functional for Spatial Structure and Orientational Ordering in Two-Dimensional Lattice Hard Rods}

\author{Paul Bitzer}
\author{Jens Weimar}
\email{jens.weimar@uni-tuebingen.de}
\author{Martin Oettel}
\affiliation{Institute of Applied Physics, University of Tübingen, Auf der Morgenstelle 10, 72076 Tübingen, Germany}

\begin{document}

\begin{abstract}
We study spatial structure and orientational ordering of horizontal and vertical hard rods on a two-dimensional square lattice using neural classical density functional theory. We construct two convolutional architectures that exploit the symmetry relating the rod orientations: a direct-$c^{(1)}$ network (learning the map between the first functional derivative of the excess free energy and the inhomogeneous density, $c^{(1)} \leftrightarrow \rho$) and a $\Phi$-network (learning directly the map between free energy and density, $F_\text{ex} \leftrightarrow \rho$). 
Both networks use exactly the same inhomogeneous grand-canonical Monte Carlo density profiles and one-body direct correlation targets.
Both neural representations substantially improve the predicted density profiles relative to the Lafuente--Cuesta lattice functional. 
The resulting free energy landscapes, however, differ between the two representations. Furthermore for the direct-$c^{(1)}$ network the integration along different density paths produces different free energies. As a result, the transition to nematic ordering for rods with aspect ratio 8 is strongly first order for the direct-$c^{(1)}$ network while only weakly discontinuous for the $\Phi$-network. The latter is a reasonable approximation to the continuous transition found in simulations, demonstrating that subtle entropic free energy differences are accessible in neural cDFT.

\end{abstract}


\section{Introduction}

Hard particles can order without attractive interactions. For rods on a square lattice, excluded volume couples two allowed orientations and produces an entropy-driven isotropic--nematic transition for rods of width 1 and length greater than $k=6$ (in lattice units).\cite{ghosh2007latticerods,matozfernandez2008latticerods} At still higher packing fractions, simulations have identified a return to a disordered state for long rods \cite{kundu2013latticerods,shah2022latticerods,vogellatticerodshighdensityestimate}. The lattice model is simple enough to permit extensive simulation, but it presents a stringent problem for classical density functional theory (cDFT): approximations for a free energy functional must describe both strongly inhomogeneous density fields and the small bulk free-energy differences that select orientational order.

In cDFT, equilibrium structure and thermodynamics follow from 
minimizing a grand potential functional, depending on the
one-body density.\cite{merminstart,bobstart} For fixed particle interactions and temperature, its intrinsic part, the free energy functional, is independent of the external potential and can therefore be reused across different external potential landscapes.
Free energy functionals from lattice fundamental measure theory (FMT), as introduced by Lafuente and Cuesta (LC), provide an established reference for hard particles on lattices.\cite{lafuente2002latticefmt,lafuentecuesta} For the present 2D rod model, however, the LC functional predicts the appearance of nematic order already at $k=4$, and for higher $k$ also at significantly lower packing fraction than simulations.
Analytical extensions of lattice LC-FMT have also been developed for square-lattice dimers \cite{michaelspaper}, but could not yet address the ordering problem of longer rods.

Machine learning aims to combine the microscopic information available from simulations with the efficiency of cDFT. Training on equilibrium density profiles allows a model to infer correlations that are difficult to capture analytically. Once trained, the functional can be used in new external fields and larger geometries, reducing the need for repeated particle simulations while providing access to both structural and thermodynamic quantities.\cite{sammuellerc1howto,reviewalessandromartin} This makes learned functionals useful for exploring confinement, interfaces, and phase behavior beyond the individual states used for training. A broad range of functional representations and network architectures has been investigated toward this goal.\cite{sammldft,sammldft2,catsroij,ml-jctc,ml-paris,kelley2024,dftmlpaircorrmatching,dftmlpaircorrmatchingddft,sammuellerpaircorrresponse,neuraloperator,feugmo2026neuralfmt,bayesml1,bayesml2}

Sammüller and co-workers introduced neural learning of the one-body direct correlation functional $c^{(1)}[\rho]$ from simulated equilibrium profiles,\cite{sammuellerc1} using the Euler--Lagrange equation to obtain training targets. Subsequent applications have described liquid--gas coexistence, fluid mixtures, ionic fluids in planar geometries and electrostatic response.\cite{sammuellerc1temperature,buiionicfluidsc1,robitschkollps,zhouc1azeotrope,buihyperdft1,buihyperdft2,buiunifiedmachinelearningframework} Neural correlation functionals have also been learned from fully resolved two- and three-dimensional profiles in single-component fluids.\cite{2dmldft,3dmldft,cheng2026equivariant} Extensions of this framework address interaction-dependent functional relations, additional equilibrium observables, and chemical-potential inference.\cite{kampa1,kampa2,kampa3,sammuellerhyperdensity,sammuellerhyperdensityhowto,sammueller2026multivariate,sammuellermu}

An internally consistent thermodynamic description of a multicomponent system requires the one-body direct correlation function $c^{(1)}_i$ of all species (labelled by $i$) to be the derivative of a common free-energy functional. In bulk, integrability requires equality of the cross derivatives of these correlations with respect to the species densities (i.e., the equality $c^{(2)}_{ij}=c^{(2)}_{ji}$ of the two-body direct correlation functions), ensuring that free-energy differences do not depend on the integration path in density space. These constraints become relevant to phase selection.
For lattice rods, horizontally and vertically oriented particles form two species and comparing isotropic and nematic (demixed) states requires consistent free-energy differences between different orientational compositions. Entropy-driven orientational ordering (with associated small free energy differences) therefore provides a rather hard test of the thermodynamic information contained in machine-learned correlations, beyond their ability to reproduce density profiles.

Here we learn free energy functionals for horizontal and vertical lattice rods of lengths $k=4$ and $k=8$, and for general density profiles, defined on the full 2D lattice. In this way, we combine advances in multidimensional learning \cite{2dmldft,3dmldft,cheng2026equivariant} with a thorough investigation of the analyticity of a learned free energy in an entropic model with subtle free energy effects.
We construct two convolutional architectures that exploit the symmetry relating the orientations. The direct-$c^{(1)}$ network (learning the map $c^{(1)} \leftrightarrow \rho$) evaluates one shared operator on symmetry-related inputs to obtain both one-body correlation fields. The $\Phi$-network (learning $F_\text{ex} \leftrightarrow \rho$) represents a local scalar free-energy density whose sum defines the excess free energy functional $F_\text{ex}$. This second architecture belongs to the broader class of learned scalar free-energy functionals.\cite{sammldft,sammldft2,catsroij,ml-jctc,ml-paris,kelley2024,dftmlpaircorrmatching,dftmlpaircorrmatchingddft,sammuellerpaircorrresponse,cheng2026equivariant,lyu2026fex}

Both networks are trained on exactly the same grand-canonical Monte Carlo (GCMC) density profiles and the same one-body direct correlation targets, with neither free energies nor bulk order parameters supplied during training. Using the LC lattice functional as an analytical reference, we assess inhomogeneous profiles, the bulk density response, free-energy landscapes, and orientational ordering. This comparison tests how integrability affects the uniqueness of the reconstructed bulk free energy and the consistency between phase equilibrium determined from $F_\text{ex}$ or directly from $c^{(1)}_i$.
The physical question is whether correlations learned solely from inhomogeneous equilibrium profiles can capture the small entropic free-energy differences governing orientational stability and phase competition in this coupled, two-dimensional system.

The paper is organized as follows.
Section~2 introduces the lattice-rod model and the density-functional framework, as well as the LC lattice functional. Section~3 presents the simulation data, common training objective, and two neural architectures. Section~4 presents the results for inhomogeneous density profiles and the bulk density response, followed by a comparison of the free-energy landscapes and orientational ordering predicted by the two neural representations. A local Landau analysis examines isotropic stability and the networks' ability to reproduce the continuous ordering transition.

\section{Lattice-Rod Model and Density-Functional Framework}

\subsection{Hard rods on a square lattice}

\begin{figure}[t]
 \centering
 \newcommand{\modelverticalrod}[3]{
  \path[modelvertical]
    ({#1+0.08},{#2+0.08}) rectangle ({#1+0.92},{#2+#3-0.08});
  \pgfmathtruncatemacro{\modeldivisions}{#3-1}
  \foreach \s in {1,...,\modeldivisions}
   \draw[modeldivision] ({#1+0.10},{#2+\s}) -- ({#1+0.90},{#2+\s});
 }
 \newcommand{\modelhorizontalrod}[3]{
  \path[modelhorizontal]
    ({#1+0.08},{#2+0.08}) rectangle ({#1+#3-0.08},{#2+0.92});
  \pgfmathtruncatemacro{\modeldivisions}{#3-1}
  \foreach \s in {1,...,\modeldivisions}
   \draw[modeldivision] ({#1+\s},{#2+0.10}) -- ({#1+\s},{#2+0.90});
 }
 \newcommand{\modelanchor}[2]{
  \draw[black,line width=0.55pt,line cap=round]
    ({#1-0.20},{#2-0.20}) -- ({#1+0.20},{#2+0.20})
    ({#1-0.20},{#2+0.20}) -- ({#1+0.20},{#2-0.20});
 }
 \resizebox{\columnwidth}{!}{
 \begin{tikzpicture}[
   x=1cm,y=1cm,font=\footnotesize,
   modelgrid/.style={draw=black!16,line width=0.18pt},
   modelvertical/.style={draw=posterhiddenborder,fill=posterhidden,
                         line width=0.35pt,rounded corners=0.35pt},
   modelhorizontal/.style={draw=posterconvborder,fill=posterconv,
                           line width=0.35pt,rounded corners=0.35pt},
   modeldivision/.style={draw=white,opacity=0.75,line width=0.22pt}
 ]
  \path[use as bounding box] (0,-0.25) rectangle (7.80,6.18);
  \node[anchor=west] at (0.10,5.99) {\textbf{(a)} Anchors, $k=4$};

  \begin{scope}[shift={(0.10,4.40)},x=0.24cm,y=0.24cm]
   \foreach \g in {0,...,5} {
    \draw[modelgrid] (\g,0) -- (\g,5);
    \draw[modelgrid] (0,\g) -- (5,\g);
   }
   \modelverticalrod{1}{1}{4}
   \modelanchor{1.5}{1.5}
  \end{scope}
  \draw[-{Latex[length=1.6mm,width=1.0mm]},black!65,line width=0.65pt]
    (1.55,5.00) -- (2.15,5.00);
  \node at (1.85,5.25) {$T$};
  \begin{scope}[shift={(2.40,4.40)},x=0.24cm,y=0.24cm]
   \foreach \g in {0,...,5} {
    \draw[modelgrid] (\g,0) -- (\g,5);
    \draw[modelgrid] (0,\g) -- (5,\g);
   }
   \modelhorizontalrod{1}{1}{4}
   \modelanchor{1.5}{1.5}
  \end{scope}

  \path[modelvertical] (4.48,5.32) rectangle (4.68,5.52);
  \node[anchor=west] at (4.87,5.42) {Vertical, $i=0$};
  \path[modelhorizontal] (4.48,4.90) rectangle (4.68,5.10);
  \node[anchor=west] at (4.87,5.00) {Horizontal, $i=1$};
  \draw[black,line width=0.55pt,line cap=round]
    (4.52,4.52) -- (4.64,4.64) (4.52,4.64) -- (4.64,4.52);
  \node[anchor=west] at (4.87,4.58) {Anchor $\mathbf r$};

  \node[anchor=west] at (0.10,3.99) {\textbf{(b)} Isotropic};
  \node[anchor=west] at (4.20,3.99) {\textbf{(c)} Nematic};

  \begin{scope}[shift={(0.10,0.30)},x=0.14375cm,y=0.14375cm]
   \foreach \g in {0,...,24} {
    \draw[modelgrid] (\g,0) -- (\g,24);
    \draw[modelgrid] (0,\g) -- (24,\g);
   }
   \foreach \rx/\ry in {1/1,7/12,14/2,21/13,3/15,10/3,18/4,13/15}
    {\modelverticalrod{\rx}{\ry}{8}}
   \foreach \rx/\ry in {3/1,15/0,0/10,12/12,4/22,14/23,8/11,12/14}
    {\modelhorizontalrod{\rx}{\ry}{8}}
   \draw[black!40,line width=0.35pt] (0,0) rectangle (24,24);
  \end{scope}

  \begin{scope}[shift={(4.20,0.30)},x=0.14375cm,y=0.14375cm]
   \foreach \g in {0,...,24} {
    \draw[modelgrid] (\g,0) -- (\g,24);
    \draw[modelgrid] (0,\g) -- (24,\g);
   }
   \foreach \rx/\ry in {1/1,12/9,0/16,23/16}
    {\modelverticalrod{\rx}{\ry}{8}}
   \foreach \rx/\ry in {0/0,8/0,16/0,3/1,14/1,2/2,12/2,4/3,16/3,10/4,3/5,13/5,6/6,15/6,2/7,12/7,5/8,16/8,0/9,15/9,3/10,14/10,13/11,1/12,16/12,0/13,14/13,3/14,13/14,2/15,16/15,2/16,15/16,1/17,13/17,3/18,15/18,5/19,2/20,14/20,6/21,14/21,1/22,13/22,3/23,15/23}
    {\modelhorizontalrod{\rx}{\ry}{8}}
   \draw[black!40,line width=0.35pt] (0,0) rectangle (24,24);
  \end{scope}
  \node[anchor=north] at (1.825,0.13) {$k=8,\quad S=0$};
  \node[anchor=north] at (5.925,0.13) {$k=8,\quad S<0$};
 \end{tikzpicture}
 }
 \caption{Schematic lattice rods and orientational order; each cell denotes one lattice site. (a) Lower/left reference sites (crosses) and the transposition $T$, illustrated for $k=4$. (b,c) Balanced and predominantly horizontal orientation populations for $k=8$, representing isotropic and nematic order, respectively.}
 \label{fig:rodmodel}
\end{figure}
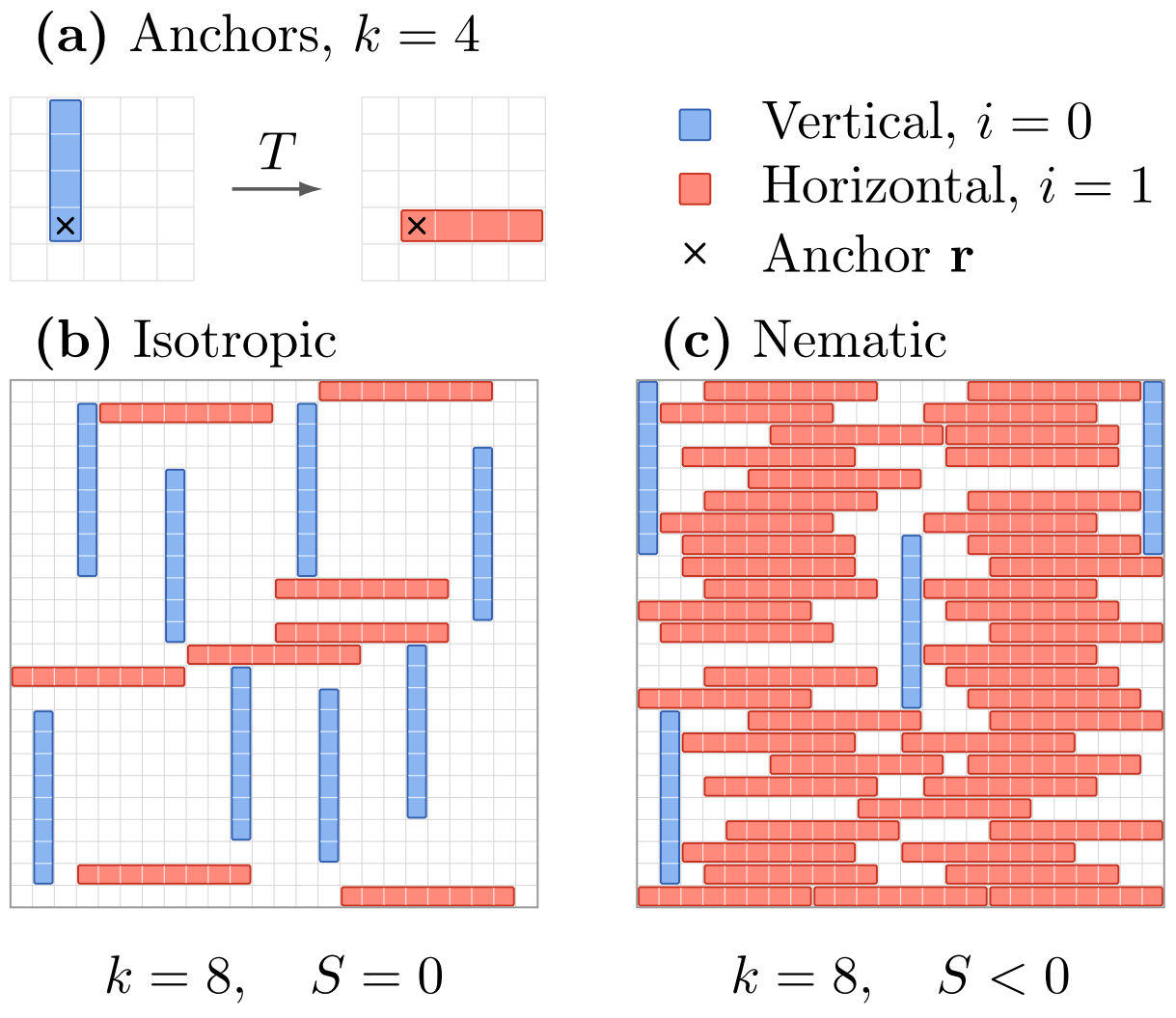

We consider a square lattice of $L_x\times L_y$ sites with periodic boundary conditions. A rod occupies $k$ consecutive sites and has one of two orientations. Species $0$ is vertical and species $1$ is horizontal. The reference position (anchor) $\mathbf r=(x,y)$ is the lower end of a vertical rod or the left end of a horizontal rod. Overlap is forbidden. The density $\rho_i(\mathbf r)$ is the mean number of rods of orientation $i$ whose reference position is $\mathbf r$. It is distinct from the local fraction of covered lattice sites. Figure~\ref{fig:rodmodel} illustrates the anchor convention and orientational ordering.

With $\rho_0$ and $\rho_1$ the average densities of vertical and horizontal rods, we define the packing fraction $\eta$ and signed orientational order parameter $S$ by
\begin{equation}
 \eta=k(\rho_0+\rho_1), \qquad
 S=\frac{\rho_0-\rho_1}{\rho_0+\rho_1}.
 \label{eq:etaS}
\end{equation}
At fixed $\eta$ the two densities are
\begin{equation}
 \rho_0(S)=\frac{\eta}{2k}(1+S), \qquad
 \rho_1(S)=\frac{\eta}{2k}(1-S).
 \label{eq:rhosS}
\end{equation}
Exchanging the coordinate axes, $(x,y)\mapsto(y,x)$, interchanges vertical and horizontal rods while preserving the lower/left anchor convention. On the density fields, this transposition acts as
\begin{equation}
 (T\rho)_i(x,y)=\rho_{1-i}(y,x),\qquad (i=0,1).
 \label{eq:speciestranspose}
\end{equation}
For homogeneous densities, $T$ exchanges $\rho_0$ and $\rho_1$ and maps $S$ to $-S$, so the bulk free-energy density is even in $S$. In a finite square box with vanishing external fields and a common chemical potential, complete sampling of this symmetry gives $\langle S\rangle=0$ even in the ordered regime. We therefore compare the magnitude of the order parameter or a selected broken-symmetry branch.

\subsection{Lattice DFT}

In a general inhomogeneous situation, lattice rods are subject to an orientation-dependent external potential $V_i(\mathbf r)$ and share a chemical potential $\mu$. We combine them into the dimensionless field $u_i(\mathbf r)=\beta[V_i(\mathbf r)-\mu]$. For trial density fields $\rho=(\rho_0,\rho_1)$, the dimensionless grand-potential functional is
\begin{equation}
 \begin{aligned}
 \beta\Omega[\rho]
 &=\sum_{\mathbf r,i}\rho_i(\mathbf r) \,
   \big(\ln\rho_i(\mathbf r)-1+u_i(\mathbf r)\big)\\
 &\quad+\beta F_{\mathrm{ex}}[\rho].
 \end{aligned}
 \label{eq:omega}
\end{equation}
The one-body direct correlation function is the negative functional derivative of the excess free energy, which reads
\begin{equation}
 c^{(1)}_{i}(\mathbf r;[\rho])
 =-\frac{\partial\,\beta F_{\mathrm{ex}}[\rho]}
         {\partial\rho_i(\mathbf r)},
 \label{eq:c1def}
\end{equation}
where we have used that, on the discrete lattice, functional differentiation reduces to ordinary partial differentiation with respect to the site densities $\rho_i(\mathbf r)$.
Stationarity of Eq.~\ref{eq:omega} gives the Euler--Lagrange equation
\begin{equation}
 \rho_i(\mathbf r)
 =\exp\!\big(-u_i(\mathbf r)+c^{(1)}_{i}(\mathbf r;[\rho])\big).
 \label{eq:EL}
\end{equation}
We use Eq.~\ref{eq:EL} both to extract training targets from equilibrium simulations and to calculate density profiles self-consistently from a learned functional.

The analytic reference is the LC lattice FMT functional.\cite{lafuente2002latticefmt,lafuentecuesta,oettel2016hardrodmonolayers} For the present two-orientation system it can be written in terms of the zero-dimensional cavity free energy $\Phi^{0\mathrm D}(n)=(1-n)\ln(1-n)+n$ as
\begin{equation}
 \begin{aligned}
 \beta F_{\mathrm{ex}}^{\mathrm{LC}}[\rho]
 &=\sum_{\mathbf r}\bigl[
   \Phi^{0\mathrm D}(n_0(\mathbf r))
   -\Phi^{0\mathrm D}(n_1^{(0)}(\mathbf r))\\
 &\hspace{5.5em}-\Phi^{0\mathrm D}(n_1^{(1)}(\mathbf r))\bigr],
 \end{aligned}
 \label{eq:LC}
\end{equation}
where
\begin{align}
 n_0(x,y)
 &=\sum_{s=0}^{k-1}\rho_0(x,y-s)
  +\sum_{s=0}^{k-1}\rho_1(x-s,y),\nonumber\\
 n_1^{(0)}(x,y)&=\sum_{s=1}^{k-1}\rho_0(x,y-s),\nonumber\\
 n_1^{(1)}(x,y)&=\sum_{s=1}^{k-1}\rho_1(x-s,y).
 \label{eq:LCweights}
\end{align}
For homogeneous rods this functional has an isotropic--nematic instability (onset of demixing between horizontal and vertical rods) at $\eta=2/(k-1)$ for $k\geq4$.\cite{oettel2016hardrodmonolayers} In contrast, in simulations the instability occurs only for $k \ge 7$, with an associated packing fraction of approximately $5/k$. For the particular case $k=8$ (investigated) here, the LC instability at $\eta=2/7 \approx 0.29$ is well below the instability found in simulations at $\eta\approx 0.64 \text{--} 0.65$ \cite{ghosh2007latticerods,matozfernandez2008latticerods,vogellatticerodshighdensityestimate}.

\section{Learning from Inhomogeneous Equilibrium Profiles}

We construct two neural representations of the excess correlations entering the Euler--Lagrange equation~\ref{eq:EL}. One, which we call the direct-$c^{(1)}$ network, maps density fields to the one-body direct correlations. The other one, which we call the $\Phi$-network, represents a local excess free-energy density $\phi$. Summation over lattice sites defines the excess free energy $F_\mathrm{ex}$, whose derivatives give the $c^{(1)}$ fields according to Eq.~\ref{eq:c1def}. Both representations are trained solely on $c^{(1)}$ targets inferred from equilibrium density profiles, following the established local learning strategy.\cite{sammuellerc1,sammuellerc1howto}

\subsection{Simulation data and common training objective}

The input density profiles are generated by grand-canonical Monte Carlo (GCMC) simulations on periodic $100\times100$ lattices for rods of length $k=8$ and $50 \times 50$ for rods of length $k=4$. Spatially varying, orientation-dependent external fields are constructed from randomized sinusoidal contributions and finite or hard plateaus (see Appendix~\ref{app:recordedmethods}). The fields for the two orientations are varied independently, while the chemical potential is common. Insertion and deletion moves generate equilibrium samples of the two density fields.

For each simulated field, rearranging Eq.~\ref{eq:EL} gives the corresponding training targets:
\begin{equation}
 c^{(1)}_{i,\mathrm{target}}(\mathbf r)
 =\ln\rho_i^{\mathrm{GCMC}}(\mathbf r)+u_i(\mathbf r).
 \label{eq:c1target}
\end{equation}
Each simulated profile thus supplies both a density input and its corresponding $c^{(1)}$ target. Entries at forbidden reference sites do not yield finite targets and are omitted from the loss. Local windows of the two density fields serve as network inputs. For either representation, the training objective is the mean-squared error in the predicted direct correlations,
\begin{equation}
\mathcal L_{c^{(1)}}
 =\left\langle
   \left|c^{(1)}_{i,\mathrm{pred}}-c^{(1)}_{i,\mathrm{target}}\right|^2
  \right\rangle,
 \label{eq:c1loss}
\end{equation}
where $c^{(1)}_{i,\mathrm{pred}}$ denotes the prediction of the direct or scalar model. The average is over sampled training windows, with contributions from masked sites with $u_i(\mathbf{r}) = \infty$ omitted. Free energies and orientational order parameters serve as predictions evaluated after training.

The $k=8$ dataset contains 872 training profiles and 100 independently generated validation profiles and  the $k=4$ dataset contains 3492 and 100 profiles respectively. The generation protocol is specified in detail in Appendix~\ref{app:recordedmethods}.

\subsection{Direct learning of the one-body correlations}

For the direct representation, we use the orientation-exchange symmetry in Eq.~\ref{eq:speciestranspose} to obtain the two correlation fields from a single network $\mathcal N_{\mathrm{dir}}$ with one output channel. The network is fully convolutional: a finite-range spatial convolution combines neighborhoods of the two density fields into local feature fields, and subsequent $1\times1$ convolutions mix the feature channels pointwise.

Evaluation on the original input $\rho$ gives $c^{(1)}_{0,\mathrm{dir}}$. To obtain $c^{(1)}_{1,\mathrm{dir}}$, we exchange the species and transpose the coordinates using $T$, evaluate the same network, and transpose the output field back:
\begin{equation}
 \begin{aligned}
 c^{(1)}_{0,\mathrm{dir}}(x,y;[\rho])
 &=\mathcal N_{\mathrm{dir}}[\rho](x,y),\\
 c^{(1)}_{1,\mathrm{dir}}(x,y;[\rho])
 &=\mathcal N_{\mathrm{dir}}[T\rho](y,x).
 \end{aligned}
 \label{eq:directsymmetry}
\end{equation}
Both evaluations share all network parameters, enforcing the orientation-exchange relation for every input density field, including fields with unequal orientation populations. Related mixture implementations use separate networks for the species-resolved correlations\cite{buiionicfluidsc1} or a common network with two outputs and species-exchange data augmentation.\cite{robitschkollps}

\begin{figure*}[p]
 \centering
 \includegraphics[width=0.92\textwidth]{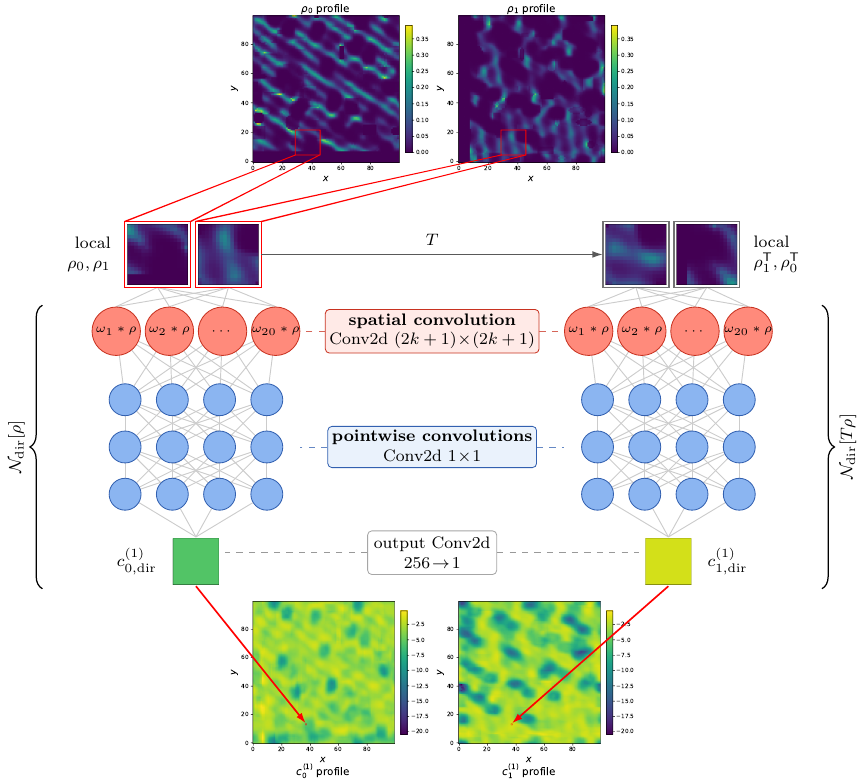}
 \caption{Fully convolutional direct-$c^{(1)}$ architecture, illustrated for $k=8$. The small full-field panels show how the same translation-equivariant operator acts on the density fields to produce the two $c^{(1)}$ fields. The operation $T$ exchanges the orientation channels of the input densities and transposes their coordinates. The finite-range $(2k+1)\times(2k+1)$ convolution maps the two density channels to 20 learned weighted-density-like feature fields and is the only layer that mixes neighboring lattice sites. Four subsequent $1\times1$ convolutions of width 256 mix channels pointwise with weights shared over the entire lattice and use SiLU activations. The nodes and connections are schematic, e.g.\ three rows represent the four pointwise layers. The left branch gives $c^{(1)}_{0,\mathrm{dir}}=\mathcal N_{\mathrm{dir}}[\rho]$. After evaluating the network on $T\rho$, the spatial coordinates of the full output field are exchanged to obtain $c^{(1)}_{1,\mathrm{dir}}(x,y)=\mathcal N_{\mathrm{dir}}[T\rho](y,x)$.}
 \label{fig:directarchitecture}
\end{figure*}

The two predictions enter Eq.~\ref{eq:c1loss} as $c^{(1)}_{i,\mathrm{pred}}=c^{(1)}_{i,\mathrm{dir}}$. Figure~\ref{fig:directarchitecture} shows the convolutional architecture and its two symmetry-related evaluations.
We also implemented a model with a regularizer, that penalizes deviations between the model's two mixed derivatives, which is called the regularized direct-$c^{(1)}$ model and is described in Appendix~\ref{app:regularizer}.

\begin{figure*}[p]
 \centering
 \includegraphics[width=0.88\textwidth]{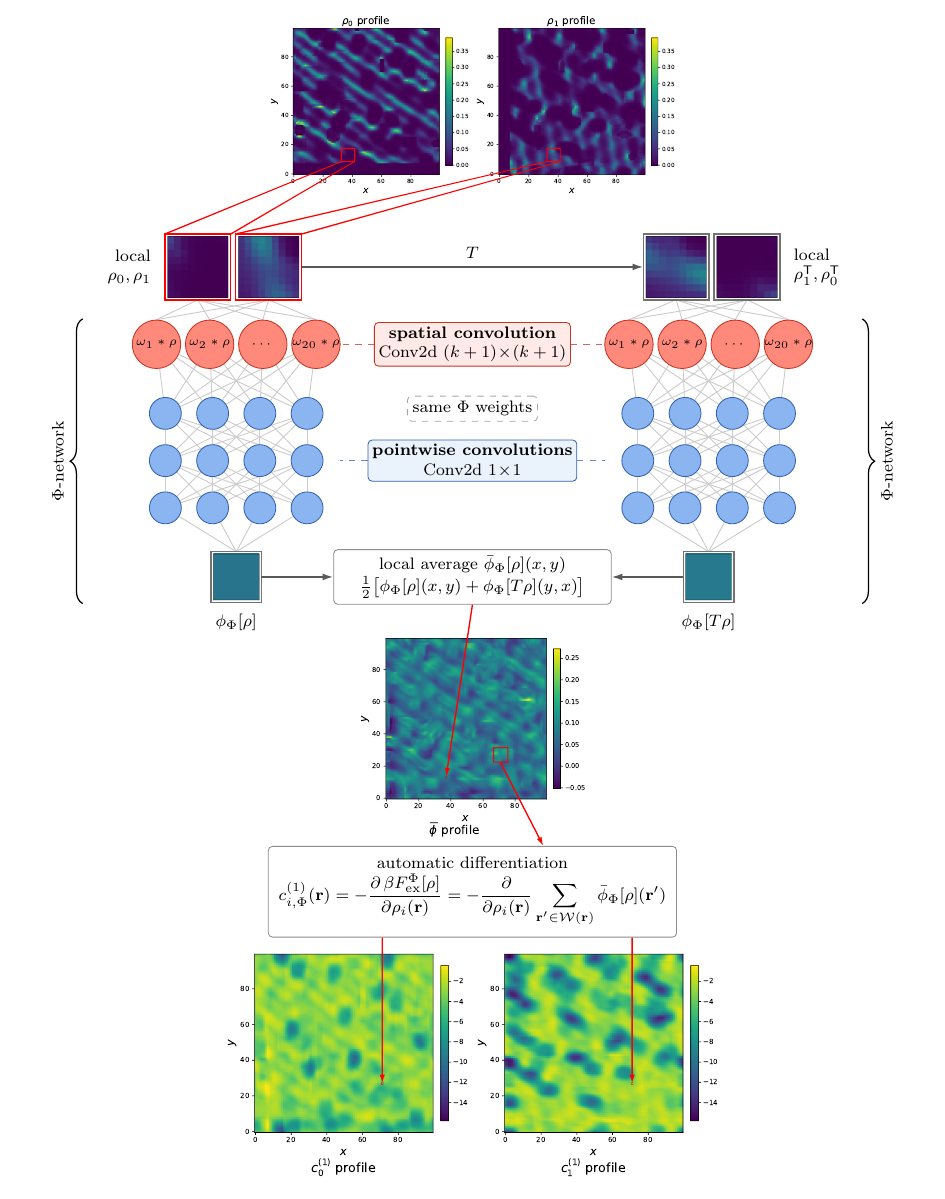}
 \caption{Scalar $\Phi$-network, illustrated for $k=8$. The two branches evaluate one shared network on $\rho$ and $T\rho$. After evaluating the network on $T\rho$, the spatial coordinates of the full output field are exchanged before averaging. The small red square in the $\phi$ map marks the local value associated with the input windows. The larger square marks the $(k+1)\times(k+1)$ neighborhood $\mathcal W(\mathbf r)$ whose local values contribute to the derivatives at the marked sites in the two $c^{(1)}$ profiles. Summation over the full scalar map gives $\beta F_{\mathrm{ex}}^\Phi$. The spatial convolution produces 20 feature channels. Four hidden $1\times1$ layers of width 256 with SiLU activations and a scalar output layer complete the network. Three blue rows schematically represent the four hidden layers. Each local scalar value depends on a $(k+1)\times(k+1)$ density window, and each $c^{(1)}$ value on a $(2k+1)\times(2k+1)$ window.}
 \label{fig:phiarchitecture}
\end{figure*}

\subsection{Learning a scalar excess free energy}

The $\Phi$-network uses a similar sequence of spatial and pointwise convolutions to represent a local scalar excess free-energy density. Summing the local contributions defines the excess free-energy functional,
\begin{align}
 \beta F_{\mathrm{ex}}^{\Phi}[\rho]
 &=\sum_{\mathbf r}\phi_{\Phi}
   \bigl(\{n_{\nu}(\mathbf r)\}\bigr),
 \label{eq:Fphi}\\
 n_{\nu}(\mathbf r)
 &=\sum_i\left(\omega_{\nu}^{(i)}*\rho_i\right)(\mathbf r).
 \label{eq:weighted}
\end{align}
Here, $n_\nu$ denotes a learned weighted density, obtained by convolving the two input density fields with finite-support kernels $\omega_{\nu}^{(i)}$. Pointwise nonlinear layers map these weighted densities to the scalar $\phi_\Phi$. This construction follows the weighted-density structure of FMT introduced in Eq.~\ref{eq:LC}, with both the convolution kernels and their nonlinear combination learned from the $c^{(1)}$ targets.

Automatic differentiation of the summed scalar output gives the predicted one-body correlations,
\begin{equation}
 c^{(1)}_{i,\Phi}(\mathbf r;[\rho])
 =-\frac{\partial\,\beta F_{\mathrm{ex}}^{\Phi}[\rho]}
         {\partial\rho_i(\mathbf r)},
 \label{eq:c1phi}
\end{equation}
which obey the orientation-exchange relation inherited from the symmetrized functional. As in Refs.~\citenum{dftmlpaircorrmatching,dftmlpaircorrmatchingddft,sammuellerpaircorrresponse,cheng2026equivariant,lyu2026fex}, we obtain the density derivatives from a learned scalar free-energy functional. Here, these derivatives enter Eq.~\ref{eq:c1loss} as $c^{(1)}_{i,\mathrm{pred}}=c^{(1)}_{i,\Phi}$, so the $\Phi$-network is constrained by the same $c^{(1)}$ targets as the direct network.

The scalar representation also relates the responses of the two correlations to changes in the density fields. Since both correlations are derivatives of a common twice-differentiable functional, they satisfy
\begin{equation}
c^{(2)}_{ij}(\mathbf r, \mathbf r' ) =
 \frac{\partial c^{(1)}_{i}(\mathbf r)}
      {\partial\rho_j(\mathbf r')}
 =
 \frac{\partial c^{(1)}_{j}(\mathbf r')}
      {\partial\rho_i(\mathbf r)} = c^{(2)}_{ji}(\mathbf r', \mathbf r ),
 \label{eq:mixed}
\end{equation}
which is the well-known exchange symmetry for the two-body direct correlation function $c^{(2)}$.
In contrast to the direct-$c^{(1)}$ network, for the $\Phi$-network, this integrability relation follows from the architecture without additional training targets. Since the derivative loss leaves an additive free-energy constant undetermined, we choose the reference $F_{\mathrm{ex}}^{\Phi}[0]=0$.

The resulting functional $F_{\mathrm{ex}}^{\Phi}$ should also be invariant under the transformation $T$. The trained network almost satisfies this invariance. However slight asymmetries remain. Therefore, during evaluation of the network we impose the orientation-exchange symmetry before summation. The scalar network is evaluated on $\rho$ and $T\rho$ with shared parameters; the second local free-energy map is evaluated at transposed coordinates $(y,x)$, and the aligned maps are averaged. The resulting functional satisfies $F_{\mathrm{ex}}^{\Phi}[T\rho]=F_{\mathrm{ex}}^{\Phi}[\rho]$. 

Figure~\ref{fig:phiarchitecture} shows the scalar architecture and the operations that turn its local output into two one-body direct correlations.

\section{Results}

\subsection{Inhomogeneous structure}

\begin{figure*}[p]
 \centering
 \includegraphics[width=0.72\textwidth]{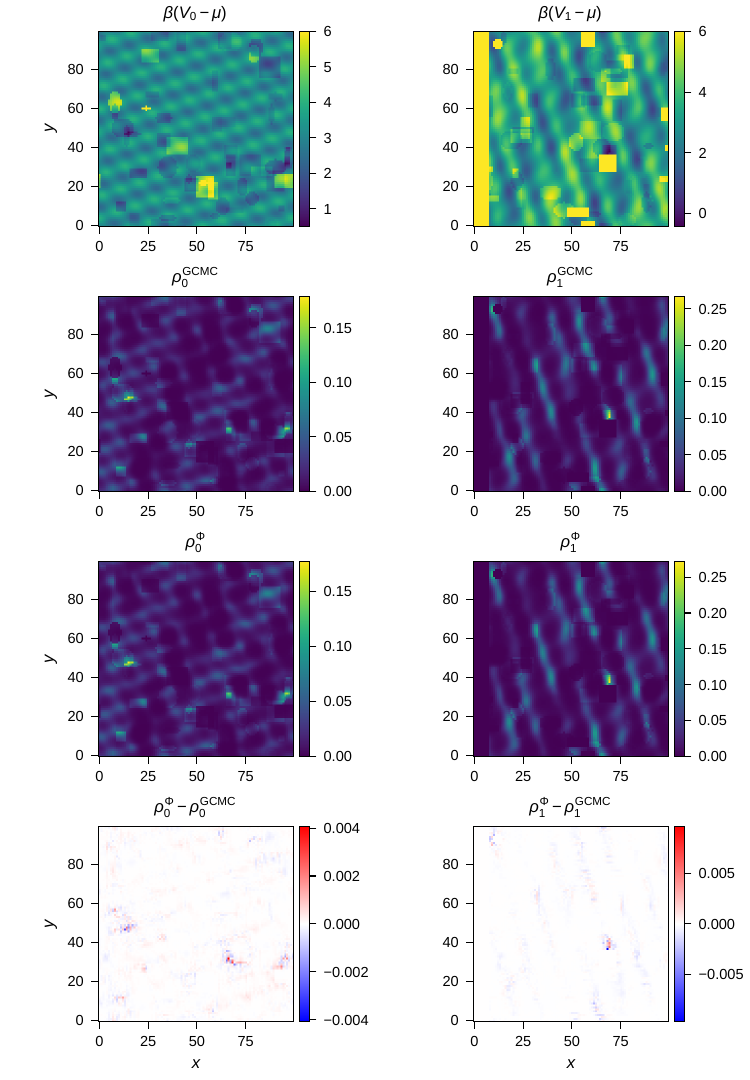}
 \caption{Representative $k=8$ validation profile for the $\Phi$-network. The columns show vertical ($i=0$) and horizontal ($i=1$) rods. From top to bottom, the rows show the effective external fields $u_i=\beta(V_i-\mu)$, the GCMC densities, the self-consistent neural densities, and their signed differences $\rho_i^\Phi-\rho_i^{\mathrm{GCMC}}$. The effective external fields are clipped at 6 for display, including at forbidden reference sites. Each color bar gives the scale for its panel.}
 \label{fig:profiles}
\end{figure*}

The detailed structural comparison focuses on rods of length $k=8$. Figure~\ref{fig:profiles} shows a representative validation profile for the $\Phi$-network. For the direct-$c^{(1)}$ network we obtain similar results. The two orientation-dependent external fields produce elongated density maxima, depleted regions, and sharp features near strongly repulsive plateaus. The self-consistent neural densities reproduce these structures in both orientations. The signed residuals are small compared with the density variation across each profile and are concentrated around some of the density maxima and their edges.

\begin{figure}[h!]
 \centering
 \includegraphics[width=\columnwidth]{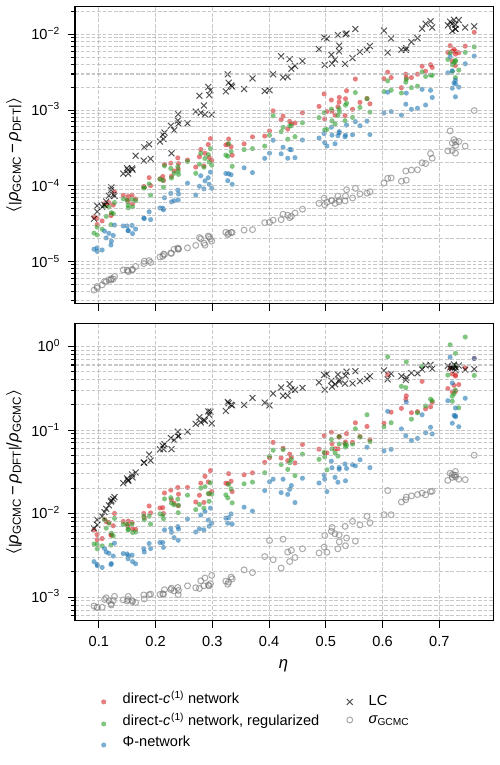}
 \caption{Errors in the $k=8$ validation profiles versus mean packing fraction. The upper and lower panels show $E_{\mathrm{abs}}$ and $E_{\mathrm{rel}}$ from Eq.~\ref{eq:profileerrors}. Blue, red, and green circles denote the $\Phi$-network, direct $c^{(1)}$-network, and regularised $c^{(1)}$-network respectively, while black crosses denote LC. Open gray circles show averages of the pointwise GCMC standard errors over the same mask, in absolute and relative forms.}
 \label{fig:profileerrors}
\end{figure}

Figure~\ref{fig:profileerrors} compares both learned representations with LC over the validation set. For each profile, the absolute and relative errors are
\begin{align}
 E_{\mathrm{abs}}&=\left\langle
   \left|\rho_i^{\mathrm{DFT}}-\rho_i^{\mathrm{GCMC}}\right|
   \right\rangle_{\mathcal M},\nonumber\\
 E_{\mathrm{rel}}&=\left\langle
   \frac{\left|\rho_i^{\mathrm{DFT}}-\rho_i^{\mathrm{GCMC}}\right|}
        {\rho_i^{\mathrm{GCMC}}}
   \right\rangle_{\mathcal M}.
 \label{eq:profileerrors}
\end{align}
Here, $\mathcal M$ contains the species specific non-zero $\rho_i(x,y)$ values with relative standard error no greater than 25\%. The mean packing fraction is calculated from the full GCMC profile. The gray symbols average the pointwise standard errors of the GCMC mean over the same mask, either directly or after division by the local GCMC density.

Both neural representations generally yield smaller errors than LC. The $\Phi$-network has the smaller absolute and relative deviations for most profiles, while the errors of all three descriptions tend to increase with packing fraction. Most neural deviations remain above the estimated GCMC uncertainty.

\subsection{Bulk density response}

\begin{figure}[h!]
 \centering
 \includegraphics[width=\columnwidth]{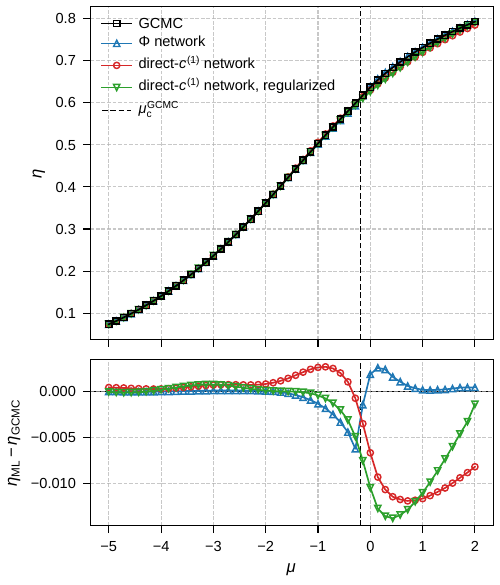}
 \caption{Equation of state for $k=8$: packing fraction $\eta$ versus $\mu$ (top) and deviations from GCMC (bottom). Symbols denote GCMC (black squares), the $\Phi$-network (blue upward triangles), direct $c^{(1)}$-network (red circles), and regularized model (green downward triangles). The dashed black line indicates the phase transition observed in GCMC simulations (see Sec.~\ref{sec:orientationalordering}).}
 \label{fig:eos}
\end{figure}

The equation of state extends the density-response test to the homogeneous system. Figure~\ref{fig:eos} compares the predicted packing fraction at fixed chemical potential with GCMC calculations without an external field and therefore $u_i(\mathbf{r}) = -\beta\mu$. Again, we also include the direct model trained with the mixed-derivative penalty defined in Appendix~\ref{app:regularizer}. The scan contains 50 chemical potentials between $\beta\mu=-5$ and $2$. All three neural models closely follow the simulated $\eta(\mu)$ up to $\mu \approx 0$ where nonmonotonic deviations set in which coincide with the isotropic-nematic instability (dashed line in Figure~\ref{fig:eos}) and which are smaller for the $\Phi$-network, compared with the others.
Figure~\ref{fig:k4eos} in the appendix also shows the equation of state for the shorter rods with $k=4$. There, no instability occurs and the deviations for increasing $\mu$ grow monotonically (on a very small level), again being smallest for the $\Phi$-network.

\begin{figure*}[tbp]
 \centering
 \includegraphics[width=\textwidth]{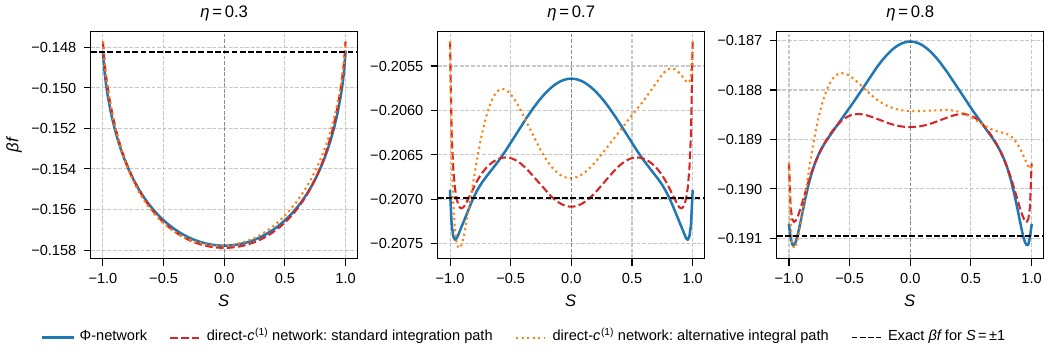}
 \caption{Bulk free-energy landscapes for $k=8$ at $\eta=0.3$, $0.7$, and $0.8$, from left to right. The ordinate $\beta f = \beta F[\rho]/(L_x \, L_y)$ denotes the dimensionless total free energy per lattice site. The solid blue curve is obtained from the $\Phi$-network. The dashed red curve reconstructs the direct-$c^{(1)}$ free energy along the simultaneous-scaling path in Eq.~\ref{eq:lineintegral}. The dotted orange curve uses the sequential path $(0,0)\to(\rho_0,0)\to(\rho_0,\rho_1)$ given in Eq.~\ref{eq:alternativepath}. All three use the reference $F_{\mathrm{ex}}[0]=0$.
 The black dashed lines give the exact $\beta f$ at $S= \pm 1$.}
 \label{fig:landscapes}
\end{figure*}

\subsection{Bulk free-energy landscapes}

After studying the response of the packing fraction to the chemical potential we now study the free energy at fixed packing fraction as a function of the order parameter. For the $\Phi$-network, which has the free energy density as its primary output, this is straightforward, as one can simply input a (homogeneous) density profile, whose values are given by Eq.~\ref{eq:rhosS}. For the direct-$c^{(1)}$ network, which only gives us the derivative of the free-energy functional, we have to integrate $c^{(1)}$.
We compare the symmetric path given by
\begin{equation}
 \beta f_{\mathrm{ex}}[\rho]- \beta f_{\mathrm{ex}}[0]
 =-\!\int_0^1\negthickspace d\alpha
   \sum_{i}\rho_i \, 
   c^{(1)}_{i}[\alpha \, \rho_0, \alpha \, \rho_1],
 \label{eq:lineintegral}
\end{equation}
which varies the two density components simultaneously and an alternative integration path that varies the species sequentially
\begin{equation} \label{eq:alternativepath}
    \begin{aligned}
        \beta f_{\mathrm{ex}}[\rho]- \beta f_{\mathrm{ex}}[0] =&-\int_0^1 \!\! d\alpha\,\rho_0 \, c^{(1)}_{0}[\alpha\, \rho_0, 0] \\
            &- \int_0^1 \!\! d\alpha\, \rho_1 \, c^{(1)}_{1}[\rho_0, \alpha\,\rho_1].
    \end{aligned}
\end{equation}
In classical DFT, the $c^{(1)}$-functionals for both components are the derivatives of a common functional and the integration path should therefore not matter. With the direct-$c^{(1)}$ network, they, however, are not and the two different integration paths can therefore deliver different results. For the $\Phi$-network, which directly calculates the free-energy density without an integration, this problem is naturally absent.

This can be seen in Fig.~\ref{fig:landscapes}, which presents the free-energy of rods of length $k=8$ as a function of the order parameter for the $\Phi$-network as well as for the direct-$c^{(1)}$ network for the two exemplary integration paths and for different packing fractions $\eta$.
For $\eta = 0.3$, all three free-energy curves are very close to each other and have a convex form with a broad minimum around $S=0$.
However, there is a small asymmetry in the curve obtained with the direct-$c^{(1)}$ network with the alternative integration path, where the minimum is not perfectly at $S=0$, but slightly shifted.
For rods of length $k=4$ the situation is very similar to this for all values of the packing fraction (not shown).
For the longer rods of $k=8$ at higher packing fractions e.g.\ $\eta=0.7$ and $\eta=0.8$, the situation is more complicated. The curves calculated with the direct-$c^{(1)}$ network via the (symmetrical) standard integration path and the $\Phi$-network are both perfectly symmetrical and have two global minima close to $\abs{S} = 1$ with rather similar values. However, beyond that, their general forms do not match. Most notably, the direct-$c^{(1)}$ network's curve has an extra local minimum around $S=0$, which is almost as deep as the two global minima. With the alternative, asymmetric (sequential) integration path, the curve is even more distinct. For both displayed packing fractions, the curve also shows a sharp global minimum at $S \to -1$ at a roughly similar value of $S$, which is however notably deeper than with the symmetric integration path. More interestingly, the slight asymmetry already reported for the lower packing fraction of $\eta = 0.3$, is very pronounced here. A local minimum for $S \to 1$ is still observable, but much less pronounced. In the case of $\eta = 0.7$ it is much less pronounced than the local minimum around $S=0$ which was already observed for the symmetric path.
The integration path given by Eq.~\ref{eq:alternativepath} uses the sequential path $(0,0)\to(\rho_0,0)\to(\rho_0,\rho_1)$. One can also switch the order and integrate along the path $(0,0)\to(0,\rho_1)\to(\rho_0,\rho_1)$. This results in the curves of Fig.~\ref{fig:landscapes} for the alternative path being mirrored along $S=0$.

For $S=\pm 1$, the LC functional gives the exact free energy (corresponding to 1D lattice hard rods), see the dashed lines in Fig.~\ref{fig:landscapes}. The $\Phi$-network almost perfectly recovers this value.
As can be seen from Fig.~\ref{fig:landscapes}, the free energy differences between isotropic and nematic states are small, on the order of $10^{-3}$, thus the integration path dependence for the direct-$c^{(1)}$ network has a significant influence on pinpointing the transition.
The $\Phi$-network, by construction, does not have this problem and is therefore a better choice for studying the system's free-energy.
Systems with attractive interactions and higher free energy differences between different phases likely would perform better here, see e.g.\ Ref.~\citenum{robitschkollps}.

\subsection{Orientational ordering} \label{sec:orientationalordering}
We compare the predicted phase behavior of the $\Phi$-network with the direct-$c^{(1)}$ network by using two different methods: via the global minima of the free-energy density $f(S)$ and the stationary states obtained from the orientational Euler--Lagrange balance.
For the direct-$c^{(1)}$ network, we choose the symmetric path (Eq.~\ref{eq:lineintegral}) as the integration path to obtain the free energy.
\begin{figure}[h!]
 \centering
 \includegraphics[width=\columnwidth]{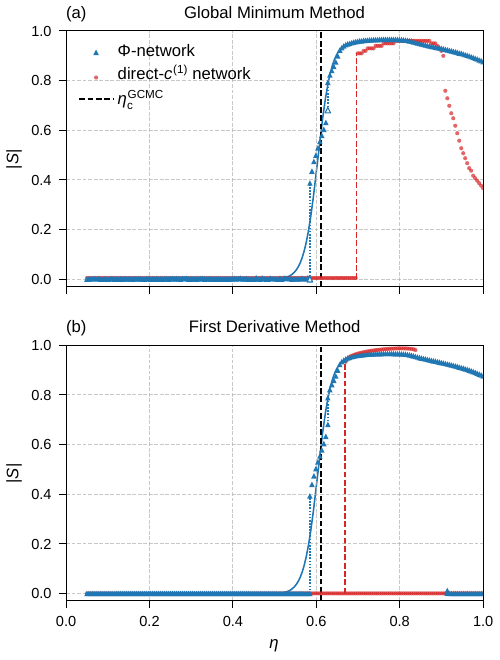}
 \caption{Bulk orientational order $|S|$ versus packing fraction $\eta$ for $k=8$. Blue triangles denote the $\Phi$-network and red circles the direct-$c^{(1)}$ network.
 The dashed black line indicates the estimate for the critical packing fraction obtained from the finite-size GCMC simulations.
 The dashed red and dotted blue lines indicate discontinuities in the phase transition.
 (a) Global minima of the sampled free-energy landscapes. The direct-$c^{(1)}$ reconstruction uses the simultaneous path in Eq.~\ref{eq:lineintegral}.
 The two empty blue triangles are borrowed from b) to better indicate the jumps and represent additional local rather then global minima calculated with that method.
 (b) Stationary branches obtained from the orientational balance in Eq.~\ref{eq:dfdS}.
 Multiple branches are retained where present.}
\label{fig:ordering}
\end{figure}
Figure ~\ref{fig:ordering}a) shows the results for $|S|(\eta)$ for the two networks.
Both show $\abs{S} = 0$ for low and intermediate values of $\eta$, which corresponds to an isotropic phase.
For the direct-$c^{(1)}$ network with the symmetric integration path, we have an abrupt, discontinuous transition to the nematic phase at $\eta \approx 0.70$, with values of $\abs{S} \approx 0.9$.

For the $\Phi$-network, orientational order emerges around $\eta \approx 0.6$.
First we observe a jump from $\abs{S} = 0$ to $\abs{S} \approx 0.4$, followed by a continuous increase in $S$ which is interrupted by a much smaller second jump between $\abs{S} \approx 0.7$ and $\abs{S} \approx 0.8$.

For high packing fractions the standard GCMC simulations quickly reach their limits, i.e.\ profiles show an increasingly high uncertainty for the mean, or, in other words, are less reliable as training data and therefore no training data for $\eta > 0.7$ was used. Therefore,  $\abs{S}(\eta)$ decreasing for packing fractions arround $0.8 \text{--} 0.9$ and above can be considered an artifact, as we are extrapolating much out of the training regime.

The black dashed line in Figure ~\ref{fig:ordering} marks the location of the maximum variance of $\abs{S}$ in our GCMC simulations, which gives a finite-size estimate for the critical packing fraction $\eta_c^\mathrm{GCMC} \approx 0.61$. This value nicely lies in the intermediate $\abs{S}$-region of the $\Phi$-network's curve (trained on the GCMC data with the same box size). Note that this critical packing fraction is of course subject to finite size errors. A careful finite size analysis in simulation studies accounting for these effects places the thermodynamic transition at $\eta_c \approx 0.64 \text{--} 0.65$. \cite{matozfernandez2008latticerods,vogellatticerodshighdensityestimate}

Stationary nematic (demixed) states can also be obtained by balancing the two orientation-specific Euler--Lagrange equations (first derivative method). For correlations generated by a common free-energy functional, this balance is the derivative of the bulk free-energy landscape:
\begin{equation}
 \begin{aligned}
 \left.\frac{\partial\,\beta f}{\partial S}\right|_{\eta}
 &=\frac{\eta}{2 \, k}\ln\left(\frac{\rho_0(S)}{\rho_1(S)}\right)\\
 &\quad-\frac{\eta}{2 \, k}
  \left(c^{(1)}_{0}[\rho(S)]-c^{(1)}_{1}[\rho(S)]\right).
 \end{aligned}
 \label{eq:dfdS}
\end{equation}
For the $\Phi$-network, this identity holds by construction. For the direct-$c^{(1)}$ network, we solve for zeros of the right-hand side using the directly learned $c^{(1)}_i$ which are not a general functional derivative of the reconstructed free energy. This avoids choosing an integration path, but the orientational balance can admit more than one symmetry-related $\pm S$-pair of stationary solutions.

In Figure~\ref{fig:ordering}b), the corresponding results for the $\Phi$-network are equal to the global minimum method except for the appearance of two extra datapoints at the previously discussed jumps and a branching, isotropic solution for $\eta > 0.9$ (representing a local minimum in the free energy).
A high-density disordered phase can be observed with advanced GCMC simulations of a homogeneous system \cite{kundu2013latticerods}, which does fit the observation $\eta \to 1$, $\abs{S}=0$. However, the appearance of this phase is clearly too early. For $k=8$, a finite-size analysis determined the transition to occur at $\eta = 0.974$. \cite{vogellatticerodshighdensityestimate}

In contrast, for the direct-$c^{(1)}$ network the results in Figure~\ref{fig:ordering}b) demonstrate that an isotropic stationary solution ($\abs{S} = 0$) is always present. A nematic stationary solution is additionally present for $\eta \approx 0.67 \text{--} 0.84$. Its disappearance beyond $\eta \approx 0.84$  is again clearly too early for the appearance of the simulation-inferred high density partially disordered phase.
Together with the results for the path-dependent free energy reconstructions, the disagreement between landscape minima and orientational balance shows a lack of internal thermodynamic consistency in the present direct-$c^{(1)}$ model.

The phases can also be obtained by employing Picard-iterations \cite{robitschkollps} to obtain density profiles for a free interface and reading off the coexisting densities from their asymptotics. In this way one basically chooses one of the branches from the first derivative method depending on the initial conditions.

\subsection{Landau analysis}
To distinguish phase selection from local isotropic stability, we use a Landau expansion of the free-energy density about $S=0$ at fixed packing fraction $\eta$:
\begin{equation}
    \begin{aligned}
        \beta f(S;\eta)
        =&\beta f(0;\eta) + \frac{1}{2} \, a_2(\eta) \, S^2 +\frac{1}{4!} \, a_4(\eta) \, S^4\\
        &+\frac{1}{6!} \, a_6(\eta) \, S^6 + O(S^{8}).
    \end{aligned}
\label{eq:landau}
\end{equation}
Here, $a_n(\eta)=\left.\partial^n(\beta f)/\partial S^n\right|_{S=0}$, with $\eta$ held fixed during differentiation. Orientation-exchange symmetry makes the odd derivatives vanish at $S=0$. For the direct-$c^{(1)}$ network we again use the symmetric integration path (Eq.~\ref{eq:lineintegral}) to calculate the free-energy density.

A second estimate of the quadratic coefficient follows from differentiating the right-hand side of Eq.~\ref{eq:dfdS} once with respect to $S$ at fixed $\eta$ and evaluating it at $S=0$:
\begin{equation}
		a_{2}(\eta) \!=\!
		 \frac{\eta}{k}\\
		 -\frac{\eta}{2k}
		\left.
		\frac{\partial}{\partial S}
		\left(\! c_0^{(1)}[\rho(S)]-c_1^{(1)}[\rho(S)]\!\right)
		\right|_{S=0} \negthickspace\negthickspace.
		\label{eq:landau_balance}
\end{equation}
For the $\Phi$-network, this is equal to $a_2(\eta)=\left.\partial^2(\beta f)/\partial S^2\right|_{S=0}$  by construction, whereas their agreement is a separate consistency test for the direct-$c^{(1)}$ network.

\begin{figure}[h!]
	\centering
	\includegraphics[width=\columnwidth]{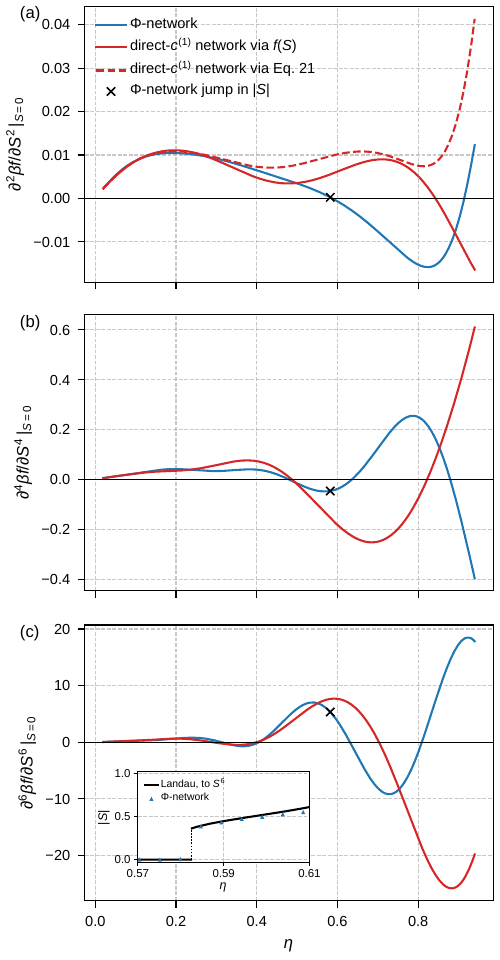}
	\caption{Local Landau analysis for $k=8$ at $S=0$. (a) Second, (b) fourth and (c) sixth derivatives of $\beta f$ with respect to $S$, taken at fixed $\eta$ and shown versus packing fraction. Blue curves denote the $\Phi$-network. Solid red curves use the symmetric direct-$c^{(1)}$ reconstruction in Eq.~\ref{eq:lineintegral}. The dashed red curve in (a) gives $a_{2}$ from Eq.~\ref{eq:landau_balance}. The odd derivatives vanish by orientation-exchange symmetry in all cases and are not shown. The inset in (c) compares the free-energy minima of the $\Phi$-network and of its sixth Landau polynomial.}
	\label{fig:landau}
\end{figure}

For the Landau free energy up to $O(S^6)$, a continuous transition emerges if $a_2$ crosses zero (the isotropic state loses local stability) while $a_4,a_6$ stay positive. A discontinuous transition may appear if $a_2>0$, $a_4<0$ and $a_6>0$ for which the system hops from the local, isotropic minimum to an ordered state with finite $|S|$.

Figure~\ref{fig:landau} shows the coefficients $a_n$. For the direct-$c^{(1)}$ network, the two different $a_2$-curves are not equal but clearly positive over the whole reliable packing fraction domain of Fig.~\ref{fig:ordering}.
The coefficient $a_4$ is negative and $a_6$ is positive across the transition region and this explains the discontinuous jump observed there.
For the $\Phi$-network, the situation is superficially more complicated. As discussed before, in Fig.~\ref{fig:ordering}, a discontinuous jump from $\abs{S} = 0$ to $\abs{S} \approx 0.4$ is observed. We highlight this point in Fig.~\ref{fig:landau}a) with a black cross. The transition happens just directly before $a_2$ changes its sign (so, $a_2 >0$ but very small), and in the vicinity of this point $a_4<0$ (small) and $a_6>0$, explaining the discontinuous jump from $S=0$ to $\approx \sqrt{-20a_4/a_6}$ (assuming $a_2=0$). (Note further that the full Landau theory up to $O(S^6)$ matches the $\Phi$-network data for $\abs{S}(\eta)$ very well around the transition, see the inset of Fig.~\ref{fig:landau}c).)
In contrast to the direct-$c^{(1)}$ network, the smallness of $a_2$ at the transition point makes the isotropic state weakly locally stable, and since also $a_4$ is small the transition can be considered a likewise weak discontinuous transition.

The small numerical values of $a_2$ and $a_4$ around the transition suggest that the network seems rather close to generating a truly continuous phase transition, but on the other hand their exact values may also depend on training procedures.

\section{Conclusions and Outlook}

For a model of 2D hard rods on a square lattice, we have explored the capabilities of neural classical DFT to determine inhomogeneous density profiles and the phase diagram. A particular challenge consisted in learning an excess free-energy functional for fully general 2D density profiles and addressing the subtle free energy differences on the order of $10^{-3}$ $k_B T$ (per lattice site) which discriminate between the isotropic and the nematic (demixed) phase.   
We have developed two convolutional neural models, the $\Phi$-network learning directly the map  $F_\text{ex} \leftrightarrow \rho$ and the direct-$c^{(1)}$ network learning the map $c^{(1)}_i \leftrightarrow \rho$ for the horizontal and vertical rod species $i$.
Both networks use exactly the same inhomogeneous GCMC density profiles and the same one-body direct correlation targets inferred from them.
They substantially improve the inhomogeneous density profiles relative to the LC functional, with the $\Phi$-network giving the smaller errors for most validation profiles. Both also reproduce the overall bulk density response to the chemical potential.
Phase diagrams can be determined by evaluating the networks at constant bulk densities without requiring free energies or bulk order parameters as training targets. For predictions based on free-energy differences between competing phases, our comparison favors the $\Phi$-network, which ensures internal thermodynamic consistency by generating both $c^{(1)}$ fields from a common functional and delivering two-body direct correlation functions $c^{(2)}$ with the correct exchange symmetry between species. Free energy determination for the direct-$c^{(1)}$ network introduces an integration path dependence whose effects are small but enough to change qualitative aspects of the entropic transition in this model.

The transition has been studied in detail for rod length $k=8$.
The $\Phi$-network captures the emergence of orientational order near $\eta \approx 0.6$ and the strongly ordered states observed in GCMC. 
The phase transition is found to be strongly discontinuous for the direct-$c^{(1)}$ network. For the $\Phi$-network a weakly discontinuous transition has been found, closer to the continuous transition expected from simulations. In a Landau analysis, the weakly discontinuous behavior could be linked to the smallness of the leading two expansion parameters.

Direct-$c^{(1)}$ learning remains an attractive choice for structural calculations when no free-energy evaluation is needed and computational speed is a priority. It supplies the required correlation fields through forward network evaluations, avoiding the differentiation of a scalar functional required by the $\Phi$-network. Generating the GCMC training data accounts for the dominant computational cost of the present approach, while training the networks is comparatively inexpensive. Once this common data set is available, both architectures can be trained at modest additional cost without further simulations. This allows both models to be kept available and the representation to be chosen for each subsequent calculation according to the required observables, accuracy, and evaluation cost.

The present models, trained on simulation data for a fixed system size, do not match the precision of dedicated simulation techniques employing e.g. finite-size scaling techniques in locating and characterizing phase transitions. Achieving that precision was not the aim of this study. Rather, we have demonstrated that for a 2D lattice model, it is possible to transfer simulation data into a numerical representation of the free energy functional without dimensional restrictions. This numerical functional (in principle) contains all equilibrium information, and we have shown that in particular the $\Phi$-network delivers inhomogeneous distributions and free energy landscapes/phase behavior at the same time, all in good agreement with simulations for this entropic system with subtle free energy effects.

It is an interesting question for the future whether it is possible to include a systematic finite-size analysis into neural DFT. Ref.~\citenum{sammuellerc1temperature} has included the system size as an input node but found no clear effects. It is perhaps necessary also to make the training architecture system size-dependent in order to capture critical correlations more properly. In this context, lattice rod models in 2D and 3D are interesting test cases with their continuous and weak first-order transitions \cite{gschwind2017,rajesh2017,quiring2019}. Furthermore, we see an extension of the neural DFT techniques to more general anisotropic models in the continuum as a worthwhile and challenging task.

\section*{Data and Code Availability}

A repository containing the code and models will be made available upon publication.

\section*{Acknowledgments}
We thank Michael Zimmermann and Alessandro Simon for useful discussions. This work is supported by the Deutsche Forschungsgemeinschaft (DFG, German Research Foundation), project 535083866. 
Generative AI (ChatGPT and Claude) assisted in coding and manuscript writing.

\appendix

\section{External Fields and Validation}
\label{app:recordedmethods}

The external potentials for the two rod orientations are generated independently from sinusoidal contributions and localized plateaus. In thermal units, they have the form
\begin{equation}
 \beta V_i(\mathbf r)=\sum_{n=1}^{N_{\mathrm{sin}}}
 A_{i,n}\sin\!\left(2\pi\mathbf r\cdot\mathbf q_{i,n}+\varphi_{i,n}\right)
 +P_i(\mathbf r),
 \label{eq:randomfield}
\end{equation}
where $N_{\mathrm{sin}}$ is drawn uniformly from $\{2,3,4,5\}$, $A_{i,n}$ from $[0.2,1]$, and $\varphi_{i,n}$ from $[0,2\pi]$. The wave vectors are $\mathbf q=(m_x/L_x,m_y/L_y)$, with integer components drawn independently from $\{-q_{\max},\ldots,q_{\max}-1\}$, where $q_{\max}=10$ for $k=8$ and 20 for $k=4$. $P_i$ denotes the contribution of another potential type which we will discuss in the following.

In addition to the periodic sinusoidal potential terms we also introduce another potential type called plateaus. A given plateau can with equal probability be either of a rectangular or an ellipsoidal shape. Each plateau is centered on a randomly chosen lattice-site, and has half-widths or semiaxes drawn independently from $[1,6]$ lattice spacings. Lattice points that lie within a given plateau have a constant value added or subtracted to their potential. Individual plateaus may overlap which can result in multiple constant values being added or subtracted to the potential at a given lattice point. The probability of a plateau being assigned a value uniformly chosen from $[0.2, 2]$ is $31.5\%$ and the probability of a plateau being assigned a value uniformly chosen from $[-1, -0.2]$ is $58.5\%$. There is also a $10 \%$ probability of a plateau being assigned a value of $+\infty$. There are 50--69 plateaus for systems generated for $k=8$ and 10--14 for $k=4$. The collective contributions of all plateaus are denoted by $P_i$ in Eq.~\ref{eq:randomfield}.
Hard strips of width eight sites are added along each of two box edges independently with probability $20 \%$.
The fields supplied to the simulations are $u_i=\beta(V_i-\mu)$. A common $\beta\mu$ is drawn uniformly from $[-5,1]$ for $k=8$ and $[-5,1.3]$ for $k=4$.
Generated training profiles which resulted in GCMC densities with packing density $\eta>0.7$ were filtered out of the training dataset. 

The inhomogeneous density profiles are obtained by Picard iteration of Eq.~\ref{eq:EL} with adaptive linear mixing. The initial density is $\rho_i=0.01$ at allowed reference sites and zero at forbidden reference sites. The iteration stops when the mean squared Euler--Lagrange residual over allowed reference sites and both orientations falls below $10^{-10}$, or after $10^4$ steps. The same procedure is used for the neural models and LC. If the stopping criterion is not reached, the last finite iterate is retained in the error comparison, so iteration errors can contribute to the reported deviations.

\begin{figure}[h!]
 \centering
 \includegraphics[width=\columnwidth]{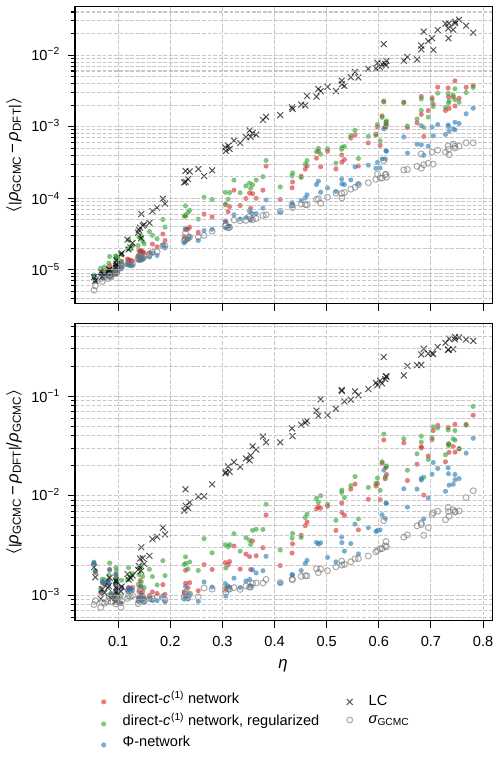}
 \caption{Validation errors for $k=4$. The upper and lower panels show $E_{\mathrm{abs}}$ and $E_{\mathrm{rel}}$ from Eq.~\ref{eq:profileerrors}. Blue, red, and green circles denote the $\Phi$-network, direct $c^{(1)}$-network, and regularized model, respectively. Black crosses denote LC. Open gray circles show the GCMC standard errors averaged over the same sites, in absolute and relative forms.}
 \label{fig:profileerrorsappendix}
\end{figure}

\begin{figure}[h!]
    \centering
    \includegraphics[width=\columnwidth]{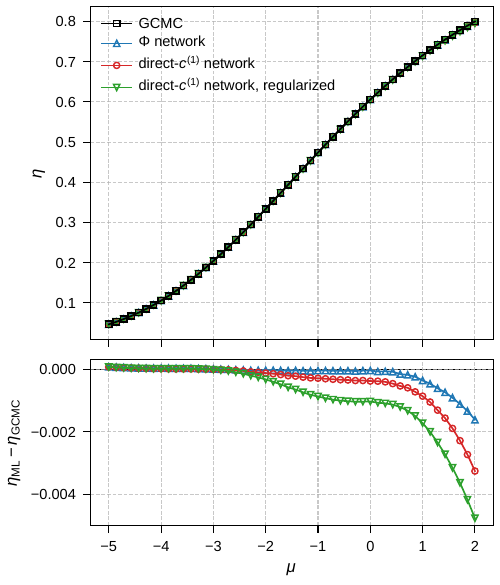}
    \caption{Equation of state for $k=4$: packing fraction $\eta$ versus $\mu$ (top) and deviations from GCMC (bottom). Symbols denote GCMC (black squares), the $\Phi$-network (blue upward triangles), direct $c^{(1)}$-network (red circles), and regularized model (green downward triangles).}
    \label{fig:k4eos}
\end{figure}

Figure~\ref{fig:profileerrorsappendix} supplements the main-text comparison with validation results for $k=4$. The errors use Eq.~\ref{eq:profileerrors}, retaining nonzero GCMC densities whose relative standard error is at most 25\%. The $\Phi$-network generally yields the smallest deviations, while the direct and regularized $c^{(1)}$ models also improve on LC over most of the displayed range.
Figure~\ref{fig:k4eos} shows the equation of state for these rods.

\section{Mixed-Derivative Penalty}
\label{app:regularizer}

The regularized direct-$c^{(1)}$ model uses the same architecture and simulation targets as the unregularized model, with an additional penalty on violations of Eq.~\ref{eq:mixed}. For a window centered at $\mathbf r_c$, let $\mathcal N_k(\mathbf r_c)$ contain the $(2k+1)^2$ sites within the convolutional range. The penalty is
\begin{equation}
 \begin{aligned}
 \mathcal L_{c^{(2)}}={}&\Bigg\langle\frac{1}{(2k+1)^2}
 \sum_{i,j=0}^{1}\sum_{\mathbf r'\in\mathcal N_k(\mathbf r_c)}
 \\
 &\quad\times\bigg[\frac{\partial c^{(1)}_{i}(\mathbf r_c)}{\partial\rho_j(\mathbf r')}
 -\frac{\partial c^{(1)}_{j}(\mathbf r')}{\partial\rho_i(\mathbf r_c)}\bigg]^2\Bigg\rangle .
 \end{aligned}
 \label{eq:c2regularizer}
\end{equation}
The brackets denote an average over training windows. Each species-pair contribution is averaged over sites, and the four contributions are summed. Automatic differentiation supplies both derivatives without simulated $c^{(2)}$ targets. Input windows of size $(4k+1)\times(4k+1)$ provide the neighborhoods required for these derivatives. The $c^{(1)}$ data loss is still evaluated only at the central site for both orientations.

\begin{figure}[htbp]
 \centering
 \includegraphics[width=\columnwidth]{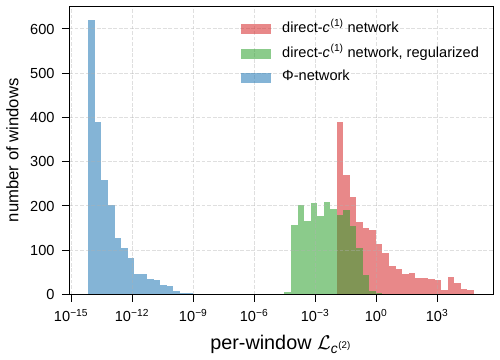}
 \caption{Histograms of the window contributions to the mixed-derivative penalty in Eq.~\ref{eq:c2regularizer} for $k=8$. Red denotes the direct-$c^{(1)}$ network, green the regularized direct model, and blue the $\Phi$-network. The horizontal axis is logarithmic.}
 \label{fig:c2error}
\end{figure}

The data loss $\mathcal{L}_{c^{(1)}}$ averages the squared errors in Eq.~\ref{eq:c1loss} over all finite species-resolved targets in the batch. The implemented objective is
\begin{equation}
 \mathcal{L}_{\mathrm{reg}}=0.8 \, \mathcal{L}_{c^{(1)}}+0.2 \, \mathcal L_{c^{(2)}}.
 \label{eq:regularizedloss}
\end{equation}
This penalty encourages reciprocal responses on sampled windows without imposing exact integrability for arbitrary density fields. The resulting profile errors and bulk density response are included in Figs.~\ref{fig:profileerrors}, \ref{fig:eos}, and~\ref{fig:profileerrorsappendix}.
As can be seen in Fig.~\ref{fig:c2error}, introducing the penalty does not fully eliminate violations of Eq.~\ref{eq:mixed}, but reduces them by several orders of magnitude. The $\Phi$-network, by design constraints, fulfills this equation perfectly up to machine precision.

\bibliography{lit}

\end{document}